\documentclass[10pt, a4paper, twocolumn, twoside]{article} 

\usepackage[english]{babel} 

\usepackage{microtype} 

\usepackage{amsmath,amsfonts,amsthm} 

\usepackage[svgnames]{xcolor} 

\usepackage[small, labelfont=bf, up]{caption} 

\usepackage{booktabs} 

\usepackage{lastpage} 

\usepackage{graphicx} 

\usepackage{enumitem} 
\setlist{noitemsep} 

\usepackage{sectsty} 
\allsectionsfont{\usefont{OT1}{phv}{b}{n}} 

\usepackage{geometry} 

\usepackage[T1]{fontenc} 
\usepackage[utf8]{inputenc} 

\usepackage{XCharter} 

\usepackage{hyperref}

\usepackage{journals}

\usepackage{fancyhdr} 
\newcommand{\shorttitle}[1]{\fancyhead[CE]{\textsl{#1}}}
\newcommand{\shortauthors}[1]{\fancyhead[CO]{\textsl{#1}}}

\fancypagestyle{firstpage}{ 
	\fancyhf{}
        \lhead{\vspace*{0.0em} \textit{\footnotesize 21$^{\rm st}$ European Workshop on White
            Dwarfs \\ Castanheira, Campos, Montgomery, eds. \\[-0.2em]
          July 23--27, 2018, Austin, Texas, USA}}
}

\date{}

\newcommand{\authorstyle}[1]{{\large\usefont{OT1}{phv}{b}{n}\color{DarkRed}#1}} 

\newcommand{\institution}[1]{{\footnotesize\usefont{OT1}{phv}{m}{sl}\color{Black}#1}} 

\usepackage{titling} 

\newcommand{\HorRule}{\color{DarkGoldenrod}\rule{\linewidth}{1pt}} 

\pretitle{
	\vspace{-30pt} 
	\HorRule\vspace{10pt} 
	\fontsize{16}{12}\usefont{OT1}{phv}{b}{n}\selectfont 
	\color{DarkRed} 
}

\posttitle{\par\vskip 10pt} 

\preauthor{} 

\postauthor{ 
	\vspace{0pt} 
	{\par\HorRule} 
	\vspace{-10pt} 
}

\newcommand{\newabstract}[1]{
    {\section*{Abstract}
    \bfseries #1}
  }

\usepackage[authoryear]{natbib}
\title{Asteroseismic cartography of hydrogen-deficient white dwarfs} 

\shorttitle{Asteroseismic cartography of hydrogen-deficient white dwarfs} 

\shortauthors{Giammichele, Charpinet, Fontaine, Brassard, Bergeron, Reindl and Baran} 

\author{
        \authorstyle{N.~Giammichele,$^1$ S.~Charpinet,$^1$ G.~Fontaine,$^2$ P.~Brassard,$^2$ P.~Bergeron,$^2$
					N.~Reindl,$^3$ and A.~S.~Baran$^4$}
	\newline\newline 
	$^1$\institution{IRAP, Université de Toulouse, CNRS, UPS, CNES, 14 avenue Edouard Belin, F-31400, Toulouse, France}\\ 
	$^2$\institution{Département de Physique, Université de Montréal,Montréal, QC H3C 3J7, Canada}\\ 
	$^3$\institution{Department of Physics and Astronomy, University of Leicester, University Road, Leicester LE1 7RH, UK}\\ 
	$^4$\institution{Uniwersytet Pedagogiczny, Obserwatorium na Suhorze, ul. Podchorażych 2, 30-084 Kraków, Polska} 
      }

\begin{document}

\maketitle 

\thispagestyle{firstpage} 


\newabstract{
  We present the results of the asteroseismic analysis of the hydrogen-deficient white dwarf PG 0112+104 from the $Kepler$-2 field. Our seismic procedure using the forward method based on physically sound, static models, includes the new core parameterization leading us to reproduce the periods of this star near the precision of the observations. This new fit outperforms current state-of-the-art standards by order of magnitudes. We precisely establish the internal structure and unravel the inner C/O stratification of its core. This opens up interesting perspectives on better constraining key processes in stellar physics such as nuclear burning, convection, and mixing, that shape this stratification over time.
  }


\section{introduction}

Hydrogen-deficient DB stars constitute approximately 20\% of the
white dwarf population. Most are believed to be produced by the so-called
born-again scenario involving a late thermal pulse that consumes the last "drops"
of hydrogen possibly remaining in the star at that pre-white-dwarf stage. Yet,
alternative channels may also contribute significantly to this population, as
proposed by \citet{reindl14,reindl15}. Exploiting pulsating DB (V777 Her)
white dwarfs with asteroseismology to probe their inner properties may therefore
reveal valuable clues about their possible formation channels.
Asteroseismology of white dwarf stars has been developed for more than three
decades, comprehensively summarized in, e.g., the reviews of \citet{fontaine08},
\citet{winget08}, and \citet{althaus10}, or more recently through the critical
discussion provided in Section 2.1 of \citet{giam17a}.
The introduction of a new approach to obtain seismic model solutions for white
dwarf pulsators that unravel their internal chemical stratification and match
better the observed period spectra is our latest improvement on this front
\citep{giam17a, giam17b}. This technique led to the first seismic cartography
of the distribution of helium, carbon, and oxygen inside the DB white dwarf
star KIC 08626021\citep{giam18}. The chemical stratification inside a white
dwarf bears the signature of all the processes that occurred during its
past history, in particular from the helium core burning phase and beyond.
Measuring it is therefore an important milestone toward understanding better
the physics of stars in their ultimate stages. In the present paper, we report
on preliminary results obtained from the seismic cartography of another pulsating
DB white dwarf, PG 0112+104.

\section{Asteroseismic analysis}

PG 0112+104 is one of the pulsating DB white dwarfs monitored continuously
with the $Kepler$-2 spacecraft during more than two months. This star exhibits
a quite rich oscillation spectrum composed of gravity modes, probing the
deep interior of the star. We exploit here the twelve extracted independent
modes, with periods ranging from 150.7 to 543.9 s (see Table 1), to perform a
complete seismic analysis to unravel the internal structure of this star.
As independent validations, we have in our possession, our most recent spectroscopic
determination of the atmospheric parameters, $T_{\rm eff}$ $=$ 31,040 $\pm$ 1,056K,
and log $g$ $=$7.83 $\pm$ 0.06 (see \citealt{bergeron95,gianninas11,tremblay13}),
as well as the estimated distance
of d $=$ 111 $\pm$ 1 pc, obtained from the Gaia parallax measurement.
We can also rely on the presence of a low-frequency spot modulation that
would correspond to a surface rotation period of 10.17404 $\pm$ 0.00067 hr,
if interpreted correctly by \citet{Hermes17}. These three entirely independent
measurements should be used to confront the soundness of solutions found for
any seismic analysis of PG 0112+104.

We rely on a forward-modeling technique using parameterized static, physically-sound stellar models, independent of stellar evolution calculations. With the help of the code LUCY \citep{Charpinet08}, the genetic algorithm designed to explore in depth the entire parameter space, we can isolate the optimal seismic model that best reproduces the oscillation properties by minimizing a merit function defined by the sum of the squared differences between theoretical and observed periods. We take advantage of our most recent developments in the definition of parameterized white dwarf models for asteroseismology \citep[e.g.,][]{giam17a,giam18}. The carbon-oxygen core profile (chemical composition and shape) is optimized simultaneously with the other parameters that define the full hydrostatic structure and the still settling helium in the envelope of the star. Let's recall that this approach has the great advantage of testing countless plausible configurations of chemical stratification as well as global parameters, to better find and assess the uniqueness of the optimal model uncovered. With this flexible technique, we can recover the optimal chemical stratification able to best reproduce the seismic observables, as hare-and-hounds experiments have demonstrated (see \citealt{giam17b}).\\

\begin{table}
	\centering
	\caption{Observed, theoretical periods, and mode identification from the optimal model obtained for PG 0112+104.}
		\begin{tabular}{ccc}
		  \hline\hline
		  Observed Per. & Theoretical Per. & Mode Id. \\
			 (s) & (s) & \\
			\hline
			159.29938 & 159.29976 & $\ell$=1, $k$=2 \\
			197.20771 & 197.20739 & $\ell$=1, $k$=3 \\
			245.72301 & 245.72298 & $\ell$=1, $k$=4 \\
			275.55884 & 275.55825 & $\ell$=1, $k$=5 \\
			319.51790 & 319.51756 & $\ell$=1, $k$=6 \\
			356.98212 & 356.98165 & $\ell$=1, $k$=7 \\
			543.95290 & 543.95210 & $\ell$=1, $k$=12\\
			\hline
			150.75042 & 150.75012 & $\ell$=2, $k$=4 \\
			168.31880 & 168.31825 & $\ell$=2, $k$=5 \\
			194.85912 & 194.85886 & $\ell$=2, $k$=6 \\
			215.34986 & 215.35002 & $\ell$=2, $k$=7 \\
			497.17680 & 497.17720 & $\ell$=2, $k$=20\\
			\hline
		\end{tabular}
		\label{table:periods}
\end{table}

\begin{table}[t]
  \caption{Preliminary estimates of some parameters derived from asteroseismology
  for PG 0112+104. When relevant, values are compared
  with available estimates from other independent methods : $T_{\rm eff}$
  (in K) and $\log g$ (in cgs) from spectroscopy, and the distance $d$ (in
  parsec) from Gaia DR2 parallax measurements.}
  \label{tab1}
  \begin{center} \begin{tabular}{ll}
                             & PG 0112+104         \\
\hline\\
    $T_{\rm eff}$ (spectro)  & 31,040 $\pm$ 1056  \\
    $T_{\rm eff}$ (astero)   & 30,787 $\pm$ 231   \\
    $\log g$ (spectro)       & 7.83   $\pm$ 0.06   \\
    $\log g$ (astero)        & 7.81   $\pm$ 0.01   \\
    $d$ (astero)             & 117.8  $\pm$ 8.2   \\
    $d$ (parallax)           & 111.0  $\pm$ 1.0    \\
\\
\hline
    \multicolumn{2}{c}{Other relevant parameters (see text)}\\
\hline\\
    Mass ($M_\odot$)         &  0.524         \\
    $\log q({\rm He})$       &  -2.33         \\
    $\log q({\rm core})$     &  -0.53         \\
    O(core)                  &  75.6          \\
\\
\hline

    \end{tabular} \end{center}

\end{table}

The seismic solution that best matches the pulsation properties of PG 0112+104 is uniquely determined around a well-defined minimum of the fitted merit function in parameter space. By inspecting Table 1, we find that the seismic fit reproduces very well the measured frequencies with a frequency dispersion of $\Delta \nu$ $=$ 0.007 $\mu$Hz, at least four orders of magnitude better than what is currently available for this star (\citep{Hermes17}).

Some parameters defining the optimal model uncovered are provided in Table 2.
This model has $T_{\rm eff}$ $=$ 30,787 $\pm$ 231 K and log $g$ $=$ 7.81 $\pm$ 0.01, which matches perfectly the independent measurements obtained from spectroscopy (within 0.25$\sigma$ for the effective temperature and 0.35$\sigma$ for the surface gravity), as can be seen in Fig.1. The second validity check that we can perform is to estimate the seismic distance obtained from the parameters of our optimal stellar model coupled with available photometry in a specific bandpass. From our results, we find an estimate of seismic distance of 117.8 $\pm$ 8.2 pc, which is in excellent agreement of the estimate of the distance from the Gaia parallax of 111 $\pm$ 1 pc.

\begin{figure}[t]
  \centerline{\includegraphics[width=1.0\columnwidth]{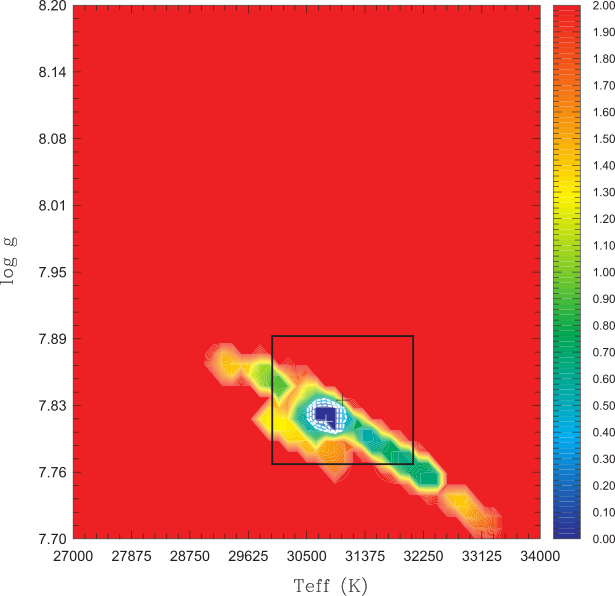}}
  \caption{Map of the merit function projected onto
the $T_{\rm eff}$-log $g$ plane for models of PG 0112+104 (on a
logarithmic scale). The white dotted curve delimits
the 1$\sigma$ confidence level relative to the best-fit
solution. The black cross surrounded by the black
box indicates the independent spectroscopic
solution and its 1$\sigma$ uncertainties.}
  \label{tefflog}
\end{figure}

The flexibility of our new parameterization coupled with an efficient search algorithm allows for a precise derivation of the internal
stratification. The oxygen mass fraction distribution in the core of the star can now be unambiguously derived, along with other
profiles for the main chemical species.
Details about the derived profiles will be given in a forthcoming paper
(Giammichele et al. 2019, in preparation). We limit the discussion here
to the findings described below.

We obtain that the central abundance of oxygen
rises up to 76\%, higher than predicted by evolutionary calculations by 10-20\%.
The extent of the homogeneous central part of the core is almost doubled, reaching a
mass fraction of log q $\approx$ -0.5, instead of log q $\approx$ -0.3. This suggests that the extent of the progenitor convective core during the core-helium-burning phase should have encompassed some 0.35 $M_{*}$ larger by a factor of 1.4 from expected. On the other hand, the behavior of the two descents in the oxygen mass stratification that bear the imprint of
helium shell burning processes from earlier stages of evolution does not differ notably from the predictions of standard
evolutionary calculations.
Moreover, we find a oxygen-carbon-helium triple transition at log q $\approx$ -2.5,
a feature predicted from evolutionary theory calculations.

The resulting helium envelope found in the optimized model is quite unexpected. PG 0112+104 only shows a rather thick layer, implying a different evolutionary path than the expected born-again scenario. The structure derived for PG 0112+104 could be
best explained if the star is instead a descendant of the very hot, helium-dominated star O(He) star, before turning into a hydrogen-deficient white dwarf, confirming the finding of \citet{reindl14}.\\

\section{Internal rotation}

Given the seismic model we have obtained for PG 0112+104, we can exploit the fine
structure uncovered by $Kepler$-2 in seven of the eleven main structures detected
in this star. Interpreted as rotational splitting, the different multiplet structures
found in the Fourier spectrum of PG 0112+104 can be used to infer the internal
rotation profile. The actual rotation profile as a function
of depth can be tested following the method developed by
\citep{Charpinet09}). In the present case, we find that the outer $\simeq$70\% of the
radius can be sounded for rotation and that PG 0112+104 rotates rigidly over this region
with a period of Prot $=$ 10.18 $\pm$ 0.27 hr.
Additionally, PG 0112+104 has a photometric signal corresponding to the surface
rotation period, inferred to correspond to 10.17404 $\pm$ 0.00067 hr. Our derived rotation
period is therefore in perfect agreement with this independent observational constraint.

\begin{figure}[t]
  \centerline{\includegraphics[width=1.0\columnwidth]{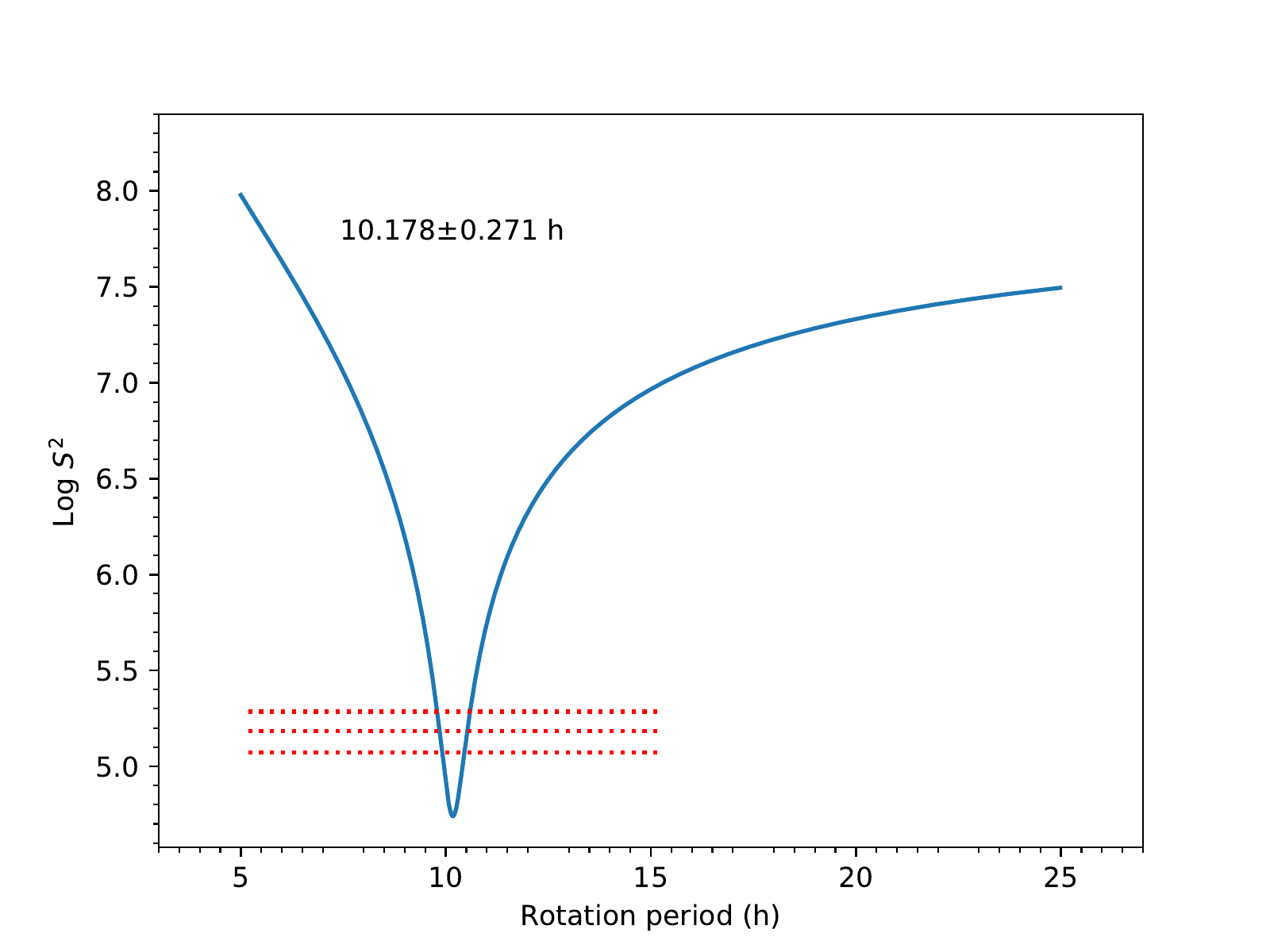}}
  \caption{Result of the optimization procedure under the hypothesis that PG 0112+104 rotates slowly and rigidly.
	This shows the behavior of the merit function in terms of the assumed rotation period.
	The merit function exhibits a very well defined minimum, corresponding to a rotation period of Prot $=$ 10.18 $\pm$ 0.27 hr.
  The dotted horizontal lines correspond, from bottom to top, to the 1$\sigma$, 2$\sigma$, and 3$\sigma$ limits.}
  \label{rot1}
\end{figure}

\begin{figure}[t]
  \centerline{\includegraphics[width=1.0\columnwidth]{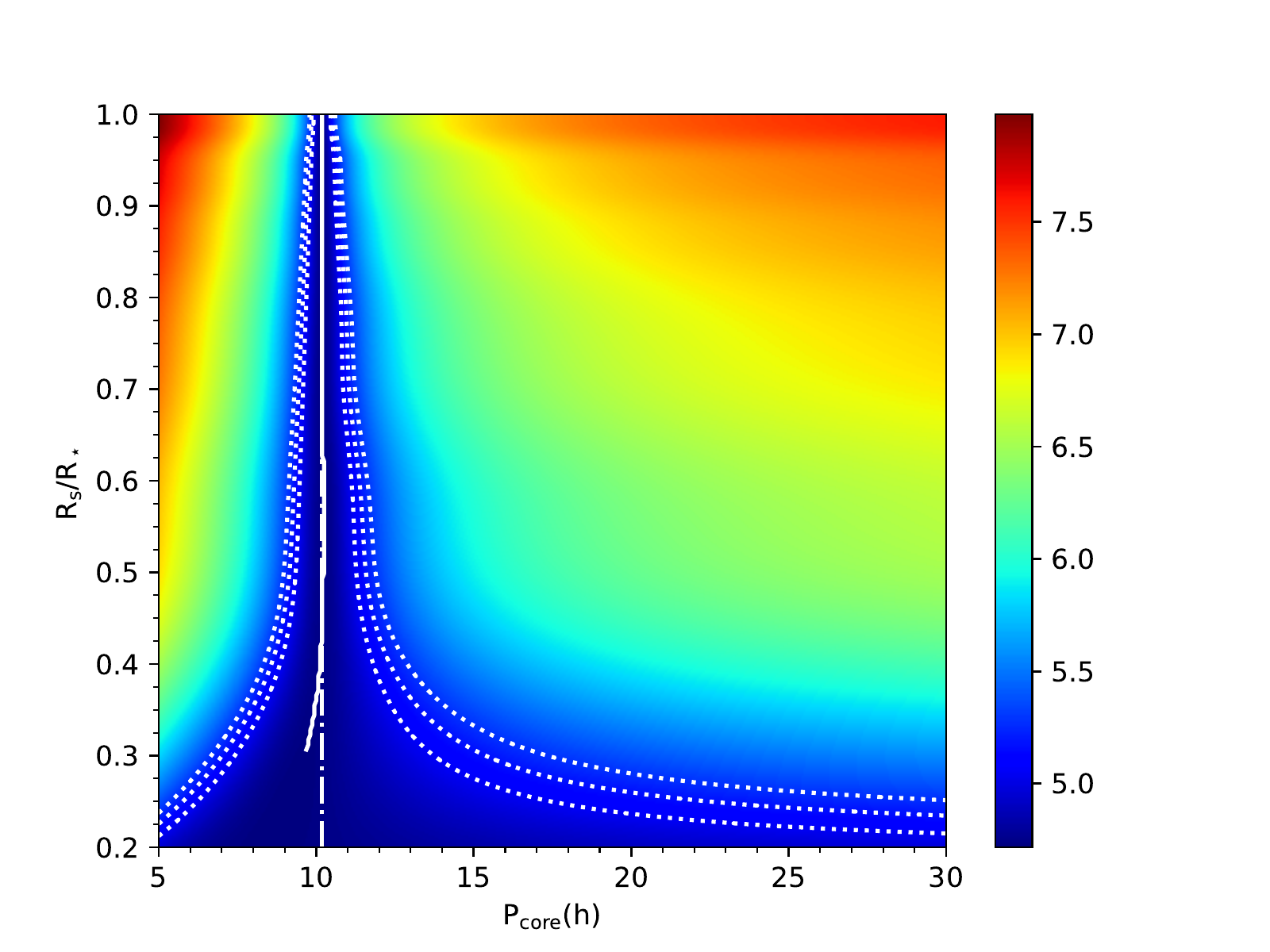}}
  \caption{Contour map of the 2D merit function that
optimizes the match between the observed spacings in
the seven multiplets with computed spacings on the basis
of our seismic model (in terms of depth and local rotation
period of the inner region in the two-zone approach of
\citet{Charpinet09}). The best-fit solution is illustrated by
the nearly vertical white curve above the solid-body
solution (vertical dot-dashed white line). The dotted white
curves on both sides of the solution depict its associated
1$\sigma$, 2$\sigma$, and 3$\sigma$ uncertainty contours.}
  \label{rot2}
\end{figure}

\vspace*{-0.5em}
\section{Conclusions}

The seismically inverted C-O stratifications in white dwarf cores offer the
opportunity to test stellar evolution models and their constitutive physics. The next
step is to explore post-AGB stellar evolution with the goal to reproduce the
seismically-derived chemical stratification. This approach is
bound to provide extremely valuable hints on which parts of the input physics
(convection, overshooting, and semi-convection treatments, nuclear physics) need
to be revised in order to conform with the seismic measurements.

In this analysis, the DB white dwarf PG 0112+104 reveals a homogeneous C/O
central core that extend further than predicted by standard evolution calculations.
An atypical structure above the core is also exposed, suggesting that this DB star may
have followed a different evolutionary path than the standard born-again scenario.
Finally, we have shown that PG 0112+104 rotates as a solid body over 70\% of
its radius with a period of 10.18$\pm$0.27 h.

\subsection*{Acknowledgment}

We acknowledge support from the Agence Nationale de la Recherche (ANR, France)
under grant ANR-17-CE31-0018, funding the INSIDE project. This work was
granted access to the high-performance computing resources of the CALMIP
computing center under allocation number 2018-p0205. G.F. acknowledges
the contribution of the Canada Research Chair Program. The authors acknowledge the Kepler
team and everyone who has contributed to making this mission possible.
Funding for the Kepler mission is provided by NASA’s Science Mission
Directorate.



\end{document}